\documentstyle[prb,preprint,aps]{revtex}
\begin{document}
\title{\hfill{\large WUB 96-22}\\[3cm]
Significance of the non-2-spinon part of $S(q,\omega)$ in the one-dimensional 
$S=1/2$ Heisenberg antiferromagnet at zero temperature}
\author{ 
	Andreas Fledderjohann and 
    Karl-Heinz M\"utter}
\address{Physics Department, University of Wuppertal, 42097 Wuppertal, Germany}
\author{Michael Karbach and 
		Gerhard M\"uller}
\address{Department of Physics, The University of Rhode Island,
Kingston RI 02881-0817}

\date{\today}
\maketitle
%
%
\begin{abstract}
%
%
The impact of the non-2-spinon excitations of the one-dimensional $S=1/2$
Heisenberg antiferromagnet on the integrated intensity, the susceptibility, the
frequency moments, and the Euclidian time representation of the $T=0$ dynamic
spin structure factor $S(q,\omega)$ is studied on the basis of finite-size data
for chains with up to $N=28$ sites.
\end{abstract}
\draft
\pacs{PACS number: 75.10 -b, 75.10.Jm, 75.40.Gb}
%
%
\section{Introduction}
%
The spectral weight in the dynamic structure factor
\begin{equation}\label{Sqw}
	S(q,\omega) \equiv \frac{1}{N} \sum_{l,n} e^{iqn}
		\int\limits_{-\infty}^{+\infty} dt e^{i\omega t} 
			\langle S_l^z(t)S_{l+n}^z\rangle
\end{equation} 
of the one-dimensional (1D) $S=1/2$ Heisenberg antiferromagnet
\begin{equation}\label{H}
	H=J\sum_{l=1}^N {\bf S}_l \cdot {\bf S}_{l+1}
\end{equation}
at $T=0$ has long been known to come predominantly from the 2-spinon
excitations,\cite{gm} which form a continuum between the lower and upper
boundaries (with energies in units of $J$)\cite{dcp,gm} 
\begin{eqnarray}
	\omega_L(q) = \frac{\pi}{2}\sin q, \quad
	\omega_U(q) = \pi \sin \frac{q}{2}.
\end{eqnarray}
However, the exact 2-spinon part of $S(q,\omega)$, here named
$S^{(2)}(q,\omega)$, was determined only very recently,\cite{boug,rhi} based on
new advances in quantum groups.\cite{JM95}

The exact expression of $S^{(2)}(q,\omega)$, which had a fairly complex
structure in its original form,\cite{boug,JM95} was simplified considerably in
Ref. \onlinecite{rhi}, evaluated numerically, and plotted versus $q$ and
$\omega$. The spectral-weight distribution was found to diverge at 
$\omega_L(q)$,
\begin{eqnarray}\label{SqtowL}
	S^{(2)}(q,\omega) & \sim & \frac{1}{\sqrt{\omega-\omega_L(q)}}
	\sqrt{\ln \frac{1}{\omega-\omega_L(q)}},
\end{eqnarray}
and to vanish in a square-root cusp at $\omega_U(q)$,
\begin{eqnarray}\label{cut}
	S^{(2)}(q,\omega) & \sim & \sqrt{\omega_U(q)-\omega}.
\end{eqnarray}
These singularities differ from those of the approximate expression
\begin{equation}\label{ma}
	S^{(a)}(q,\omega) =  
	\frac{\Theta\biglb(\omega-\omega_L(q)\bigrb)
	  \Theta\biglb(\omega_U(q)-\omega\bigrb)}{\sqrt{\omega^2-\omega_L(q)^2}}
\end{equation}
for the 2-spinon dynamic structure factor, which had been inferred from
finite-$N$ data, sum rules, and Bethe-ansatz calculations.\cite{gm} Otherwise,
the line shapes of $S^{(2)}(q,\omega)$ and $S^{(a)}(q,\omega)$ do not differ all
that much.\cite{rhi}

The 2-spinon excitations were shown to account for 82.18\% of the total 
intensity
in $S(q,\omega)$. Where in $(q,\omega)$-space is the remaining spectral weight,
from which classes of Bethe-ansatz solutions does it originate, and how does it
affect various quantities that can be derived from the dynamic structure factor?

A study of the 4-spinon contribution to $S(q,\omega)$ along the lines of
Refs. \onlinecite{JM95,boug,rhi} will shed light on the first two
questions.\cite{bkm} In answer to the third question, we investigate here the
effects of the non-2-spinon excitations on four quantities which are related to
$S(q,\omega)$ and which can be computed with high precision from finite-$N$ data
for the ground-state wave function.
%
%
\section{Integrated Intensity}
%
%
The integrated intensity of $S(q,\omega)$, i.e. the static spin structure factor
\begin{equation}\label{Iq}
	I(q) \equiv \int\limits_0^\infty \frac{d\omega}{2\pi} S(q,\omega),
\end{equation}
has been determined with high precision for wave numbers $q\leq 13\pi/14$ from
finite-$N$ data of cyclic chains with $N\leq 28$ sites.\cite{mkkhm} This 
result is plotted in Fig.~1 for comparison with the exact 2-spinon integrated
intensity $I^{(2)}(q)$ calculated from $S^{(2)}(q,\omega)$ via (\ref{Iq}) and
the integrated intensity\cite{gm}
\begin{equation}\label{Ia}
	I^{(a)}(q) = \frac{1}{2\pi}\ln \frac{1+\sin(q/2)}{\cos(q/2)}
\end{equation}
obtained from the approximate result $S^{(a)}(q,\omega)$. 

The finite-$N$ data indicate that $I(q)$ increases linearly from zero for small
$q$ and diverges logarithmically at $q=\pi$. The initial rise of the finite-$N$
data, $I(q)\to 0.271 q/\pi$,\cite{sum} is significantly steeper than that of the
2-spinon contribution, $I^{(2)}(q)\to 0.237q/\pi$ and that of the approximate
result, $I^{(a)}(q)\to 0.25q/\pi$. Hence the integrated intensity of the
non-2-spinon part of $S(q,\omega)$ increases linearly in $q$ too.

The function $I^{(a)}(q)$ approximates the 2-spinon integrated intensity
$I^{(2)}(q)$ quite well for $q/\pi\lesssim 0.6$. At larger $q$, this is no
longer the case. The divergence predicted by expression (\ref{Ia}), $I^{(a)}\sim
-\ln(1-q/\pi)$, is weaker than the divergence of the exact 2-spinon result,
$I^{(2)}(q)\sim [-\ln(1-q/\pi)]^{3/2}$.  The inset of Fig.~1 shows the relative
non-2-spinon integrated intensity, $\Delta I^{(2)}(q)=1-I^{(2)}(q)/I(q)$, and
the relative deviation, $\Delta I^{(a)}(q)=1-I^{(a)}(q)/I(q)$, of the
approximate result (\ref{Ia}). If it can be assumed that the leading singularity
of $I(q)$ at $q=\pi$ is produced entirely by the 2-spinon part of $S(q,\omega)$,
then the function $\Delta I^{(2)}(q)$ must approach zero as $q\to\pi$. The
dashed line in the inset does not rule out that this assumption is
correct. Whether or not the 4-spinon excitations contribute to the leading
singularity in $I(q)$ remains to be seen.\cite{bkm}

It is interesting to compare these results with the exact integrated intensity
of the Haldane-Shastry model,\cite{hsm} $I^{(HS)}(q) = -(1/4)\ln(1-q/\pi)$,
where non-2-spinon excitations have zero spectral weight in $S(q,\omega)$.
It turns out that for $q\lesssim 13\pi/14$, $I^{(HS)}(q)$ is a much
better approximation of $I(q)$ than $I^{(2)}(q)$ is.\cite{mkkhm}
%
%
\section{Susceptibility}
%
%
The $q$-dependent susceptibility at $T=0$ is related to $S(q,\omega)$ via the 
sum
rule\cite{hbg}
\begin{equation}\label{chiq}
	\chi(q)\equiv \frac{1}{\pi}\int\limits_0^\infty 
		\frac{d\omega}{\omega} S(q,\omega).
\end{equation}
This quantity, which has been determined with considerable accuracy from
finite-$N$ data, is plotted in Fig.~2 for comparison with the exact 2-spinon
susceptibility $\chi^{(2)}(q)$ calculated\cite{rhi} from $S^{(2)}(q,\omega)$ via
(\ref{chiq}) and the approximate result\cite{gm}
\begin{equation}\label{chiqa}
	\chi^{(a)}(q)=\frac{1}{\pi^2}\frac{q}{\sin q}
\end{equation}
inferred from (\ref{ma}).

The normalization of $S^{(a)}(q,\omega)$ was chosen such that the exact value of
the direct susceptibilty,\cite{griff} $\chi(0)=1/\pi^2$, is correctly
reproduced. With increasing $q$, $\chi^{(a)}(q)$ deviates in a downward
direction from $\chi(q)$. Its divergence at $q=\pi,~ \chi^{(q)}(q)\sim
(\pi-q)^{-1}$, is slightly weaker than that of the exact 2-spinon
susceptibilty,\cite{rhi} $\chi^{(q)}(q)\sim \sqrt{-\ln(\pi-q)}/(\pi-q)$, which,
presumably coincides with the leading singularity of $\chi(q)$.

The contribution of the non-2-spinon spectral weight of $S(q,\omega)$ to
$\chi(q)$ in the limit $q\to 0$ is small but non-negligible as indicated by the
result $\pi^2\chi^{(2)}(0)=0.9500...$, which falls some 5\% short of the exact
direct susceptibilty. The relative non-2-spinon contribution to the
susceptibility, $\Delta\chi^{(2)}(q)=1-\chi^{(2)}(q)/\chi(q)$, stays smaller for
all $q$-values than the relative non-2-spinon integrated intensity $\Delta
I^{(2)}(q)$ (see inset). This indicates that the non-2-spinon spectral weight is
located predominatly above the 2-spinon continuum.
%
%
\section{Frequency moments}
%
%
Yet a different way to assess the non-2-spinon part of $S(q,\omega)$ employs 
the frequency moments
\begin{equation}\label{fmom}
	K_n(q) \equiv \int\limits_0^\infty \frac{d\omega}{2\pi}
		\; \omega^n S(q,\omega), \quad n=1,2,...,
\end{equation}
which are related, via sum rules,\cite{sumrule} to short-range multi-spin
correlations in the ground state. For $n=1$ we know the exact
results,\cite{hbg}
\begin{eqnarray}\label{K1q}
	K_1(q) & = & \frac{2E_0}{3N}(1-\cos q),
\end{eqnarray}
where $E_0=-N(\ln 2 -1/4)$ is the ground-state energy. For $n=2,3,4,5$
high-precision results have been calculated from finite-$N$ data for the
associated ground-state expectation values.\cite{sum} The moments $K_n^{(2)}(q)$
of the exact 2-spinon dynamic structure factor $S^{(2)}(q,\omega)$ have been
determined in Ref.~\onlinecite{rhi} and the moments $K_n^{(a)}(q)$ of
$S^{(a)}(q,\omega)$ in Ref.~\onlinecite{sumrule}.

For $n=1$, both $K_1^{(2)}(q)$ and $K_1^{(a)}(q)$ reproduce the $q$-dependence
of the exact sum rule (\ref{K1q}) correctly, but the prefactors are smaller,
\begin{equation}
	\frac{K_1^{(2)}(q)}{K_1(q)} = 0.8039..., \quad
	\frac{K_1^{(a)}(q)}{K_1(q)} = 0.8462...,
\end{equation}
which again reflects the missing spectral weight of the non-2-spinon
excitations. The $q$-dependence of the moment ratios
\begin{equation}\label{Rnq}
	R_n(q) \equiv \frac{K_n(q)}{K_1(q)}, \quad \text{ for } n=2,3,4,5,
\end{equation}
of the full dynamic structure factor as inferred from finite-$N$ data are shown
in Fig.~3 along with the corresponding moment ratios $R_n^{(2)}(q)$ of
$S^{(2)}(q,\omega)$ and the moment ratios $R_n^{(a)}(q)$ of $S^{(a)}(q,\omega)$.
The most striking observation is that $R_n(q)$ approaches a nonzero value as
$q\to 0$, whereas the exact and the approximate 2-spinon moment ratios both 
go to
zero: $R_n^{(2)}(q) \sim R_n^{(a)}(q) \sim q^{n-1}$. This means that for long
wavelengths the frequency moments $K_n(q), n\geq 2$, are dominated by
non-2-spinon excitations, which are necessarily located above the narrow
2-spinon band. In other words, the 2-spinon dynamic structure factor
$S^{(2)}(q,\omega)$ does not contribute to the leading $O(q^2)$ term of $K_n(q)$
for $n\geq 2$.

At larger wave numbers, the impact of the non-2-spinon excitations on the moment
ratios is more modest but still significant. Here the deviation of
$R_n^{(2)}(q)$ from $R_n(q)$ is almost $q$-independent, and it grows with
increasing $n$. This again indicates that the non-2-spinon spectral weight comes
for the most part from higher frequencies than the 2-spinon spectral weight.

The moment ratios $R_n^{(a)}(q)$ agree very well with $R_n^{(2)}(q)$ at small
$q$, but then deviate upwardly. For $q\gtrsim \pi/3$, they rise even above the
ratios $R_n(q)$. This discrepancy, which becomes more conspicuous with
increasing $n$, reflects the fact that $S^{(a)}(q,\omega)$ underestimates the
spectral weight near the lower continuum boundary $\omega_L(q)$ and
overestimates the spectral weight near the upper boundary $\omega_U(q)$.
%
%
\section{Euclidian time representation}
%
%
In order to study the significance of the non-2-spinon excitations in a
dynamical quantity by the same kind of comparison, it is useful to consider the
Laplace transform of the dynamic structure factor,\cite{rec}
\begin{eqnarray}\label{Eucl}
	\tilde{S}(q,\tau) & \equiv & \int\limits_0^\infty
	\frac{d\omega}{2\pi} \; e^{-\omega\tau} S(q,\omega),
\end{eqnarray}
which can be interpreted as a Euclidian time representation of $S(q,\omega)$. 
For
$\tau=0$, this is the integrated intensity (\ref{Iq}). From finite-$N$ data for
$S(q,\omega)$ as obtained via recursion method for systems with $N\leq 28$
sites,\cite{rec} this quantity can be accurately extrapolated to $N\to\infty$ if
$q\neq \pi$. For the graphical representation, it is convenient to plot the
function
\begin{equation}\label{rhoqt}
	\rho(q,t)\equiv\frac{\tilde{S}(q,0)}{\tilde{S}(q,\tau)} -1, \quad
	t=\sqrt{\omega_1(p)\tau}\exp[\omega_1(p)\tau],
\end{equation}
which was used for the finite-$N$ extrapolation, instead of $\tilde{S}(q,\tau)$
itself.  The resulting curves, shown as solid lines in Fig.~4 for several
$q$-values, rise from zero with zero initial slope and then become almost
linear in $t$ with a $q$-dependent slope.

If the threshold singularity were a square-root divergence as predicted by
(\ref{ma}), then the asymptotic growth of $\rho(q,t)$ would be exactly
linear. The logarithmic correction in the exact 2-spinon threshold singularity
(\ref{SqtowL}), however, leads to a slight modification of the asymptotic growth
of $\rho(q,t)$,
\begin{eqnarray}
	\rho^{(2)}(q,t) & \stackrel{t\to\infty}{\longrightarrow} & 
			\propto \frac{t}{\ln\ln t}.
\end{eqnarray}

The dashed lines in Fig.~4 show the function $\rho^{(2)}(q,t)$ as inferred from
the exact 2-spinon dynamic structure factor. The discrepancies are fairly small
over the range of $t$ shown. The deviation
$\Delta^{(2)}(q,t)=\rho^{(2)}(q,t)-\rho(q,t)$ is shown in the inset. Not
surprisingly, the function $\rho^{(a)}(q,t)$ inferred from (\ref{ma}) deviates
more strongly from $\rho(q,t)$. This comparison was already made in
Ref.\onlinecite{rec}.
%
\acknowledgements
%
The work at URI was supported by NSF Grant DMR-93-12252, and by the Max Kade
Foundation. 
%
%

%
%
%
%
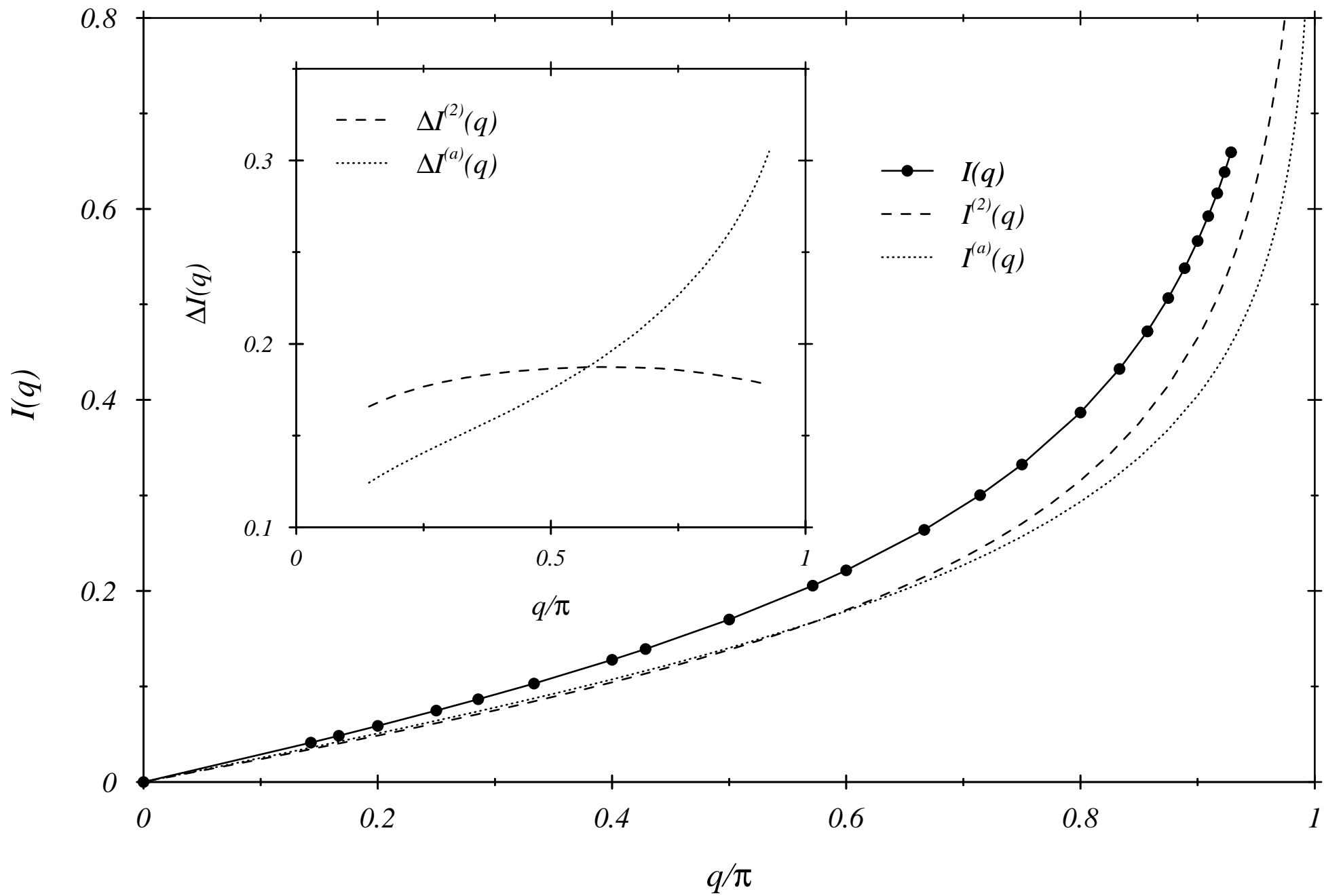
\begin{figure}
\caption[1]{
	Integrated intensities (\ref{Iq}) in comparison: $I(q)$ is represented 
        by finite-$N$ data for $N=6,8,...,28$. $I^{(2)}(q)$ is the exact 
        2-spinon result. $I^{(a)}(q)$ is the approximate result (\ref{Ia}). 
        The inset shows
 	$\Delta I^{(2)}(q)=1-I^{(2)}(q)/I(q)$ and
	$\Delta I^{(a)}(q)=1-I^{(a)}(q)/I(q)$.}
\end{figure}

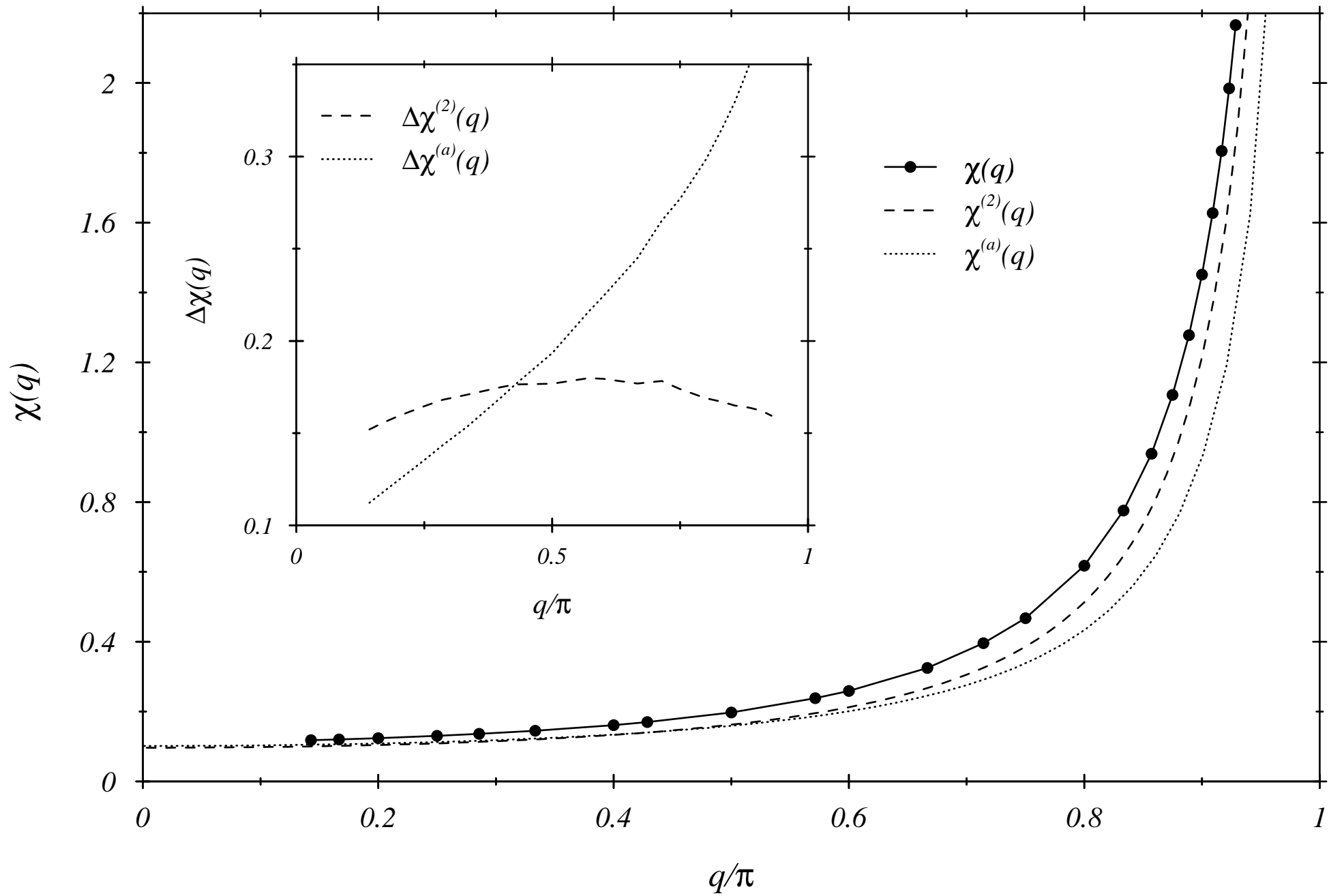
\begin{figure}
\caption[2]{ The $q$-dependent susceptibilty (\ref{chiq}) in comparison:
	$\chi(q)$ is represented by finite-$N$ data for
	$N=6,8,...,28$. $\chi^{(2)}(q)$ is the exact 2-spinon
	result. $\chi^{(a)}(q)$ is the approximate result (\ref{chiqa}). The 
        inset shows the relative non-2-spinon integrated intensity 
	$\Delta\chi^{(2)}(q)=1-\chi^{(2)}(q)/\chi(q)$ and
	$\Delta\chi^{(a)}(q)=1-\chi^{(a)}(q)/\chi(q)$.}
\end{figure}

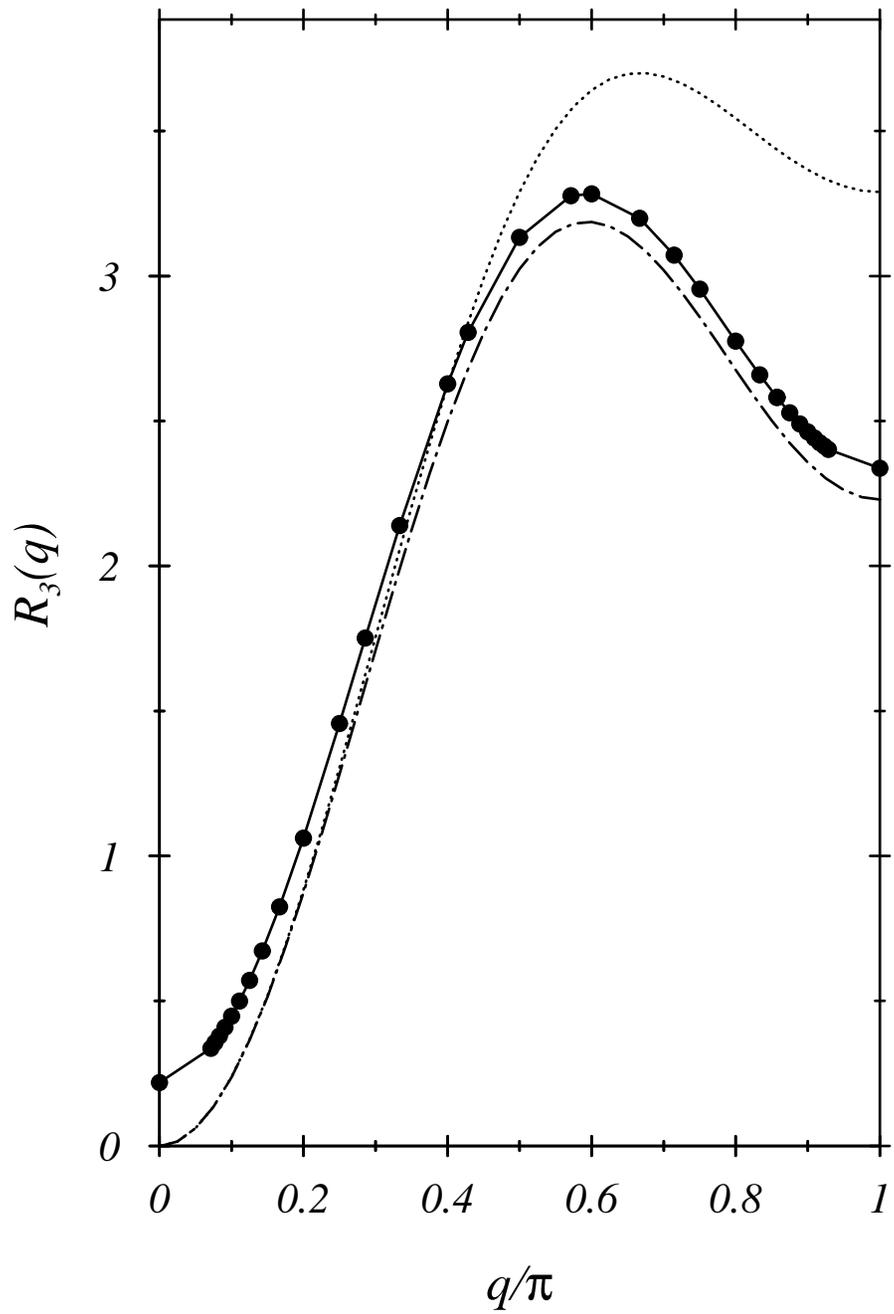
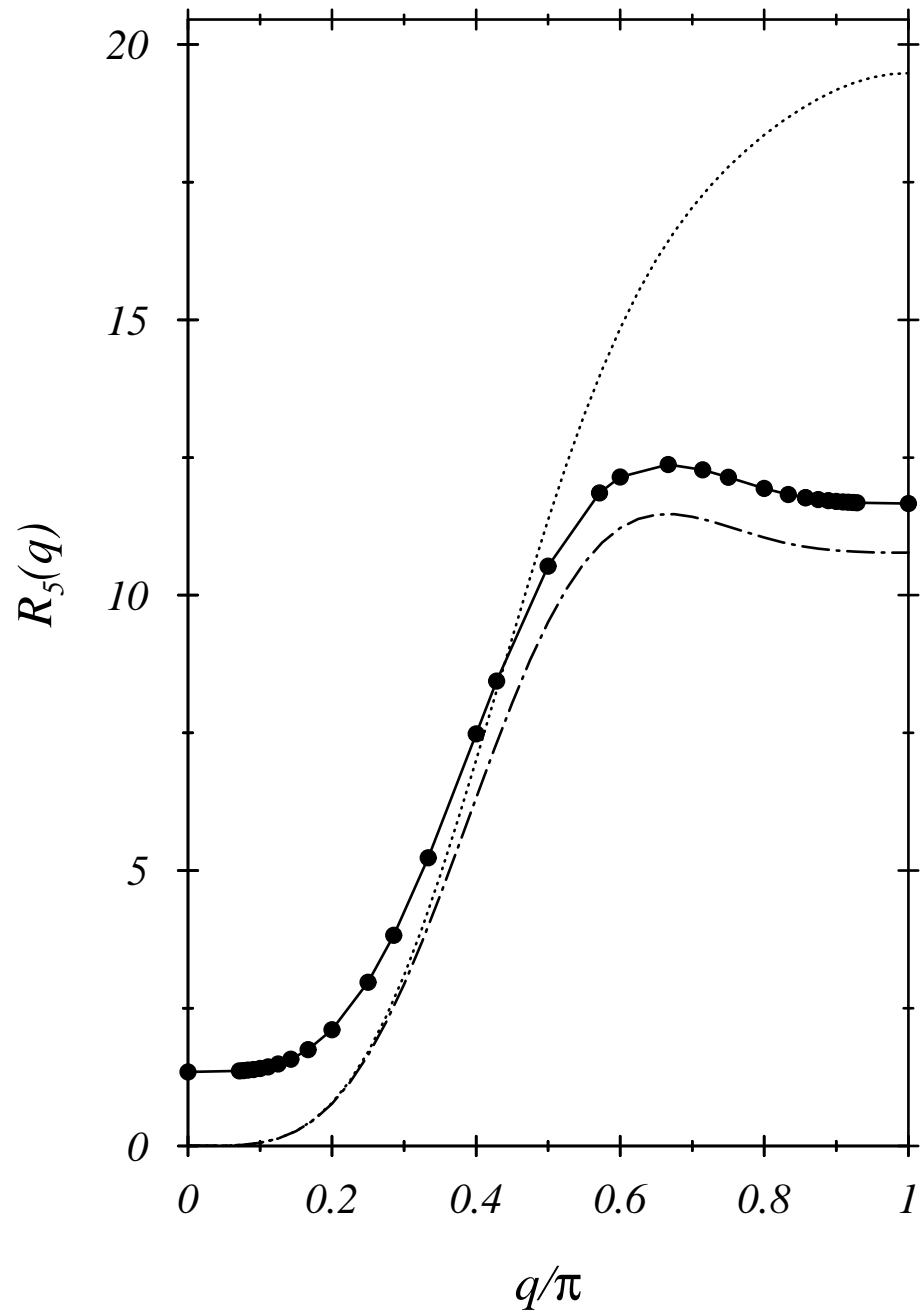
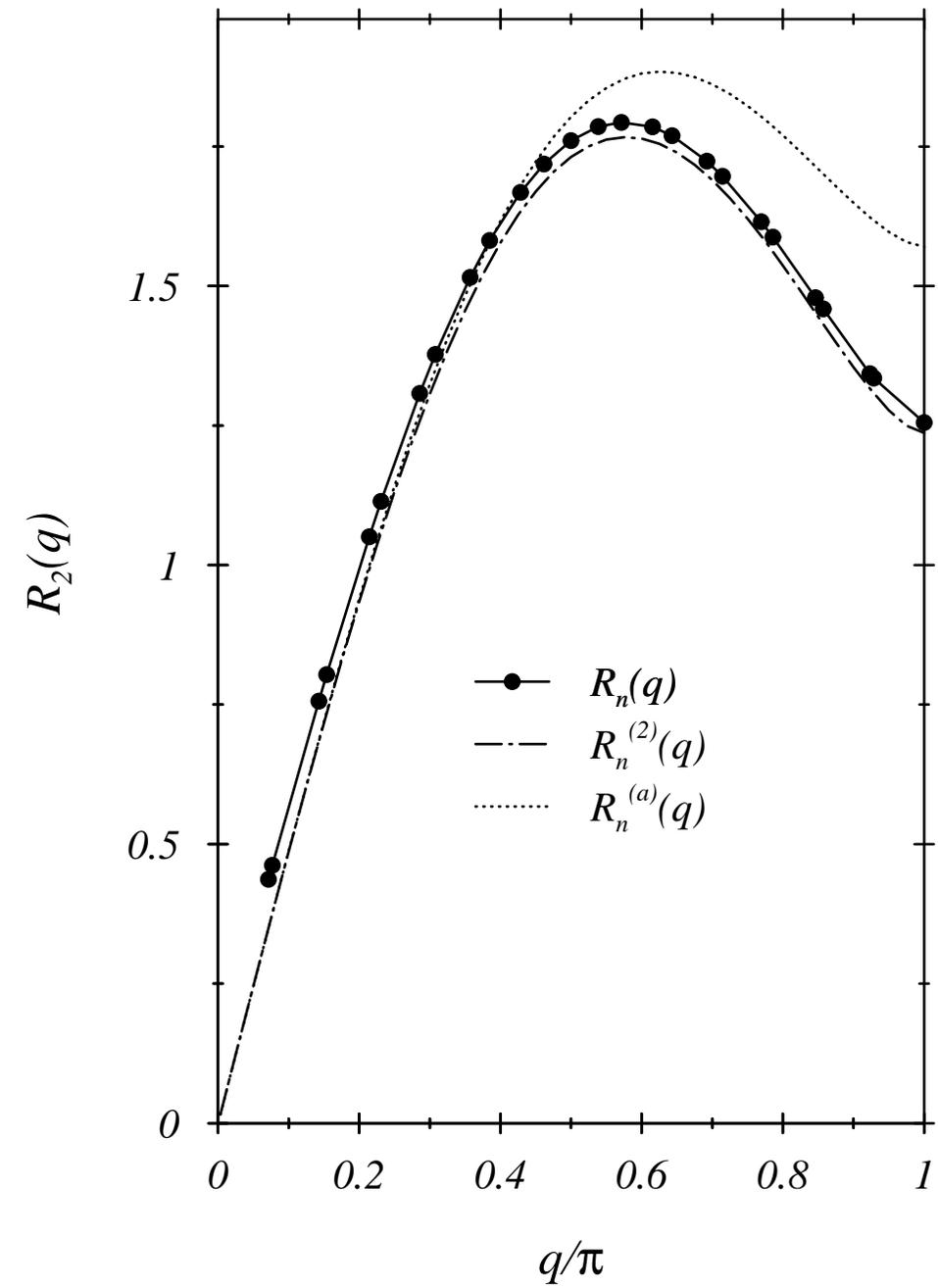
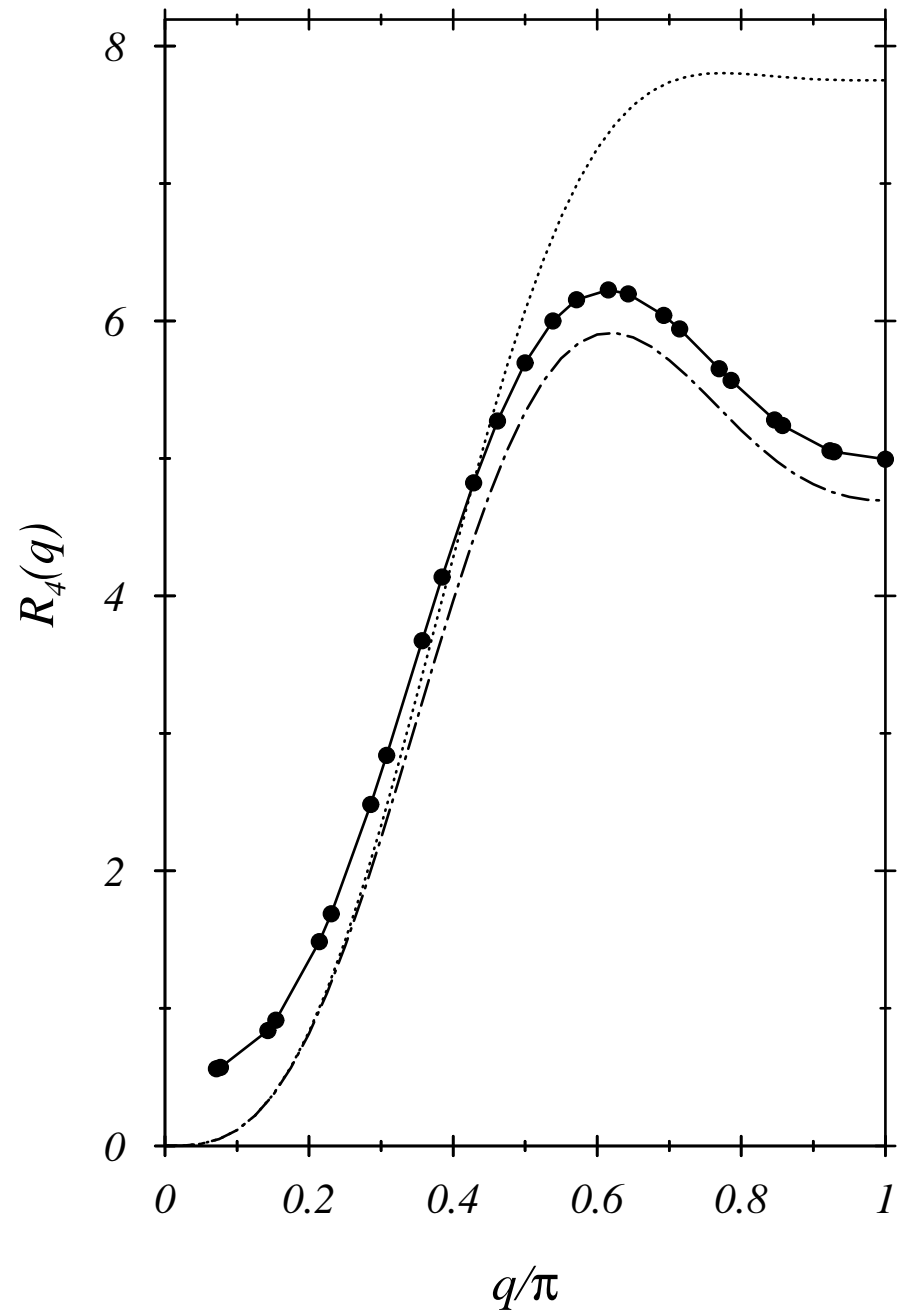
\begin{figure}
\caption[3]{ Ratios of frequency moments (\ref{Rnq}). For $n=2,3,4,5$, the
	$R_n(q)$ represent finite-$N$ data for $N=6,...,28$. The $R_n^{(2)}(q)$
        are exact 2-spinon results and the $R_n^{(a)}(q)$ are the moment ratios
        for (\ref{ma}).}
\end{figure}

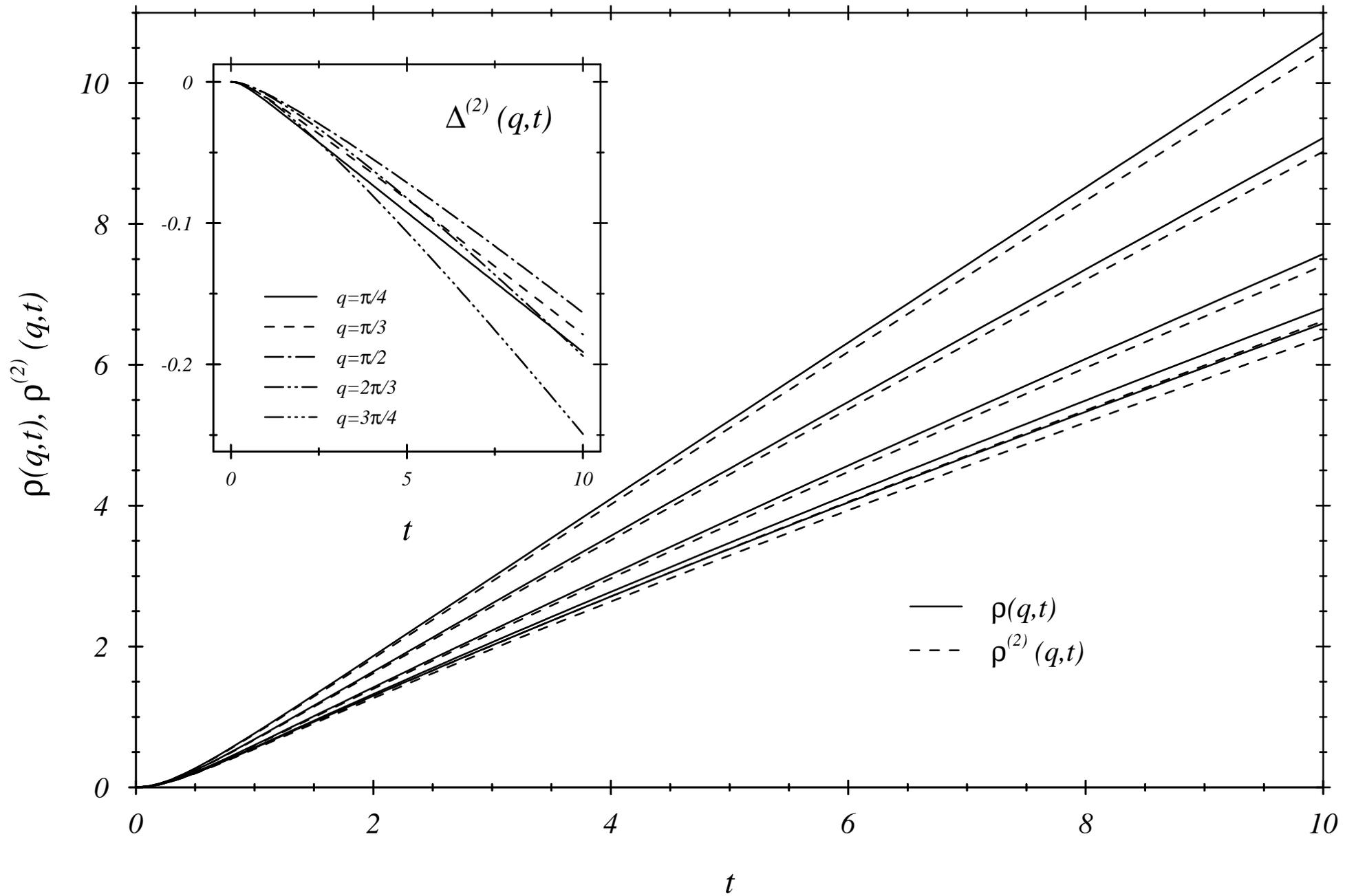
\begin{figure}
\caption[4]{The function (\ref{rhoqt}) for $q=\pi/4$, $\pi/3$, $\pi/2$, 
        $2\pi/3$, $3\pi/4$. (bottom to top) The solid lines represent 
        extrapolated finite-$N$ data and the dashed lines the exact
	2-spinon-part of that function. The inset shows the non-2-spinon part
	$\Delta^{(2)}(q,t)$ of the same function.}
\end{figure}
\end{document}